# A Novel Chaotic Encryption Scheme based on Pseudorandom Bit Padding

Sodeif Ahadpour, Yaser Sadra and Zahra ArastehFard

Department of Sciences, University of Mohaghegh Ardabili, Ardabil, IRAN.

Ahadpour@uma.ac.ir, Sadra@staff.um.ac.ir
z.arastehfard@msn.com

## Abstract

Cryptography is always very important in data origin authentications, entity authentication, data integrity and confidentiality. In recent years, a variety of chaotic cryptographic schemes have been proposed. These schemes have typical structure which performed the permutation and the diffusion stages, alternatively. The random number generators are intransitive in cryptographic schemes and be used in the diffusion functions of the image encryption for diffused pixels of plain image. In this paper, we propose a chaotic encryption scheme based on pseudorandom bit padding that the bits be generated by a novel logistic pseudorandom image algorithm. To evaluate the security of the cipher image of this scheme, the key space analysis, the correlation of two adjacent pixels and differential attack were performed. This scheme tries to improve the problem of failure of encryption such as small key space and level of security.

## 1. Introduction

The Cryptography is always very important in data origin authentications, entity authentication, data integrity and confidentiality [1-6,28,29,30]. In recent years, the cryptographic schemes have suggested some new and efficient ways to develop secure image encryption [1]. These schemes have typical structure which performed the permutation and the diffusion stages alternatively. However, most of algorithms be faced with some problems such as the lack of robustness and security. The random number generators are intransitive in cryptography for generation of cryptographic keys, allegorically, secret keys utilized in symmetric cryptosystems [2,3] and large numbers is intransitive in asymmetric cryptosystems [4,6], because of unpredictable, should better be generated randomly. In addition, random number generators in many cryptographic protocols, such as to create challenges, blinding value are used [7,8,9]. Also, the random number generators are used more in the diffusion functions of the image encryption for diffused pixels of plain image.

Random number generators can be classified into three classes which are pseudorandom number generators (PRNGs), true random number generators (TRNGs) and hybrid random number generators (HRNGs). PRNGs use deterministic processes to generate a series of outputs from an initial seed state [10,11,12]. TRNGs use of non-deterministic source (i.e., the entropy source), along with some processing function (i.e., the entropy distillation process) to generate the random bit sequence [2]. These sources consist of physical phenomena such as atmospheric noise, thermal noise, radioactive decay and even coin-tossing [13]. Many PRNGs using chaotic maps have been established. Most of them have very complex structures. In this paper, we propose a chaotic encryption scheme based on pseudorandom bit padding that the bits be generated by a novel logistic pseudorandom image algorithm. The random bit sequences produced by this generator are evaluated using the 15 statistical tests recommended by U.S. NIST [2]. Experimental results show that this PRNG possess good uniformity and randomness properties.

This paper is arranged as follows. In section 2, the properties of the logistic map are discussed. In section 3, we introduce the proposed random number generators and then discuss the uniformity and randomness of the bit sequences generated by the Proposed PRNG. In section 4, we propose chaotic encryption scheme based on pseudorandom bit padding and finally, in Section 5, we conclude the paper.

## 2. The logistic map

The logistic map is one of the most studied discrete chaotic maps. It is well-known as very sensitive to both system variable and control parameter. In addition, other features such as ergodicity, pseudo-randomness and unpredictable behavior. Therefore, it possesses great potential for various cryptographic applications such as image encryption [15,16], public key cryptography [17], key agreement protocol [9], block cipher [3], and hash function [18,21,22]. It was first proposed as pseudo random number generator by Von Neumann in 1947 partly because it had a known algebraic distribution and mentioned later, in 1969, by Knuth [23,24]. The simplest form of the logistic map is given by:

$$x_{n+1} = rx_n(1-x_n)$$

Where $x_n \in (0,1)$ and $r$ are the system variable and control parameter, respectively, and $n$ is the number of iterations. Thus, given a control parameter $r$ and a system value $x_0$; time series of logistic map $\{x_n\}_{n=0}^{\infty}$ is computed. Here, we refer to $x_0$ and $r$ as the initial state of the logistic map. In the following we use the chaotic logistic map for cryptographic applications, as follows:

$$x_{n+1} = rx_n(1-x_n) \quad (1)$$
$$x_n \in (0,1), \quad \text{and} \quad r_x \in (3.99996, 4]$$

[25]. As stated in [26], The choice of $r$ in the equation above guarantees the existence of a chaotic orbit that can be shadowed by only one map as stated in . In addition, the above map is supposed to have good qualities as a PRNG when $r \cong 4$ [25].

## 3. The proposed PRNG and randomness analysis

### 3.1 The proposed PRNG using logistic pseudorandom image algorithm

Subheadings In this section, we introduce a proposed pseudorandom number generator based on the logistic pseudorandom image algorithm. For cryptographic purposes, the output of RNGs needs to be unpredictable [2]. In this method, we use a black white dynamic image because we can use each pixel as a key. On the other hand, key space of the PRNG is a black white dynamic image. To consider a gray scale image with the size of $2^k \times 2^k$ (here, $2^8 \times 2^8$) pixels (see Fig.1(a)).

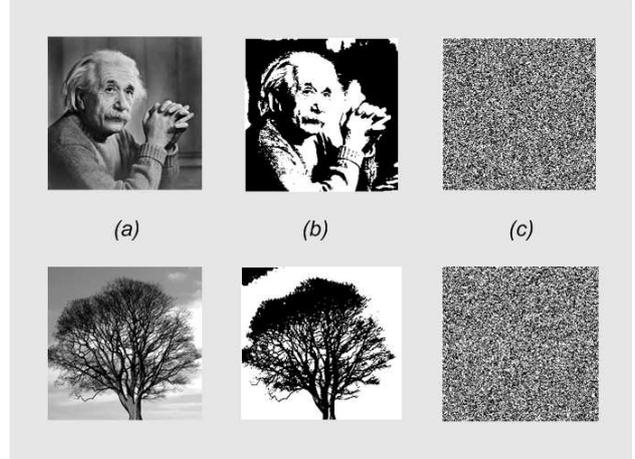

Fig. 1 (a) two gray scale images with the size of $2^8 \times 2^8$ pixels, (b) two black white images that the color of all pixels that are smaller than the Average Pixel Intensity (API) with black and all pixels that are greater than or equal to the API with white are changed, (c) two black white dynamic images that are the perfect seeds for PRNG.

We redefined it as a matrix $C_{2^k \times 2^k}$. This matrix is composed of color of the pixels in the uint8 (output range 0 to 255). Uint8 is a MATLAB built-in function. Matrix components corresponding to image pixels can be showed as $c_{ij}$. Then, we can get Average Pixel Intensity (API) [5,14]. Hence,

$$API = \frac{\sum_{i=1}^{2^k} \sum_{j=1}^{2^k} c_{ij}}{2^k \times 2^k}, \quad (2)$$

then, we change the color of all pixels that are smaller than the Average Pixel Intensity(API) with black and all pixels that are greater than or equal to the API with white (see Fig.1(b)). Now, using a two-dimensional chaotic system which is defined as follows:

$$x_n = f(x_{n-1}) \quad n = 0,1,2,...$$
$$y_n = g(y_{n-1}) \quad n = 0,1,2,...$$

that $f$ and $g : I \to I$ ($I = [0,1]$) are nonlinear maps, we get coordinates of a point $(x_n, y_n)$ in two-dimensional space. Using the following transformation can be converted coordinates of a point $(x_n, y_n)$ in two-dimensional continuous space into a point $(u(x_n), u(y_n))$ in two-dimensional discrete space of the image matrix components:

$u : I \to N \quad I = [0,1], \quad N = [1, 2^k]$

$$(i,j) = \begin{cases} u(x_n) = [x_n \times (2^k - 1)] + 1 \\ u(y_n) = [y_n \times (2^k - 1)] + 1 \end{cases} \quad (3)$$

where symbol of [ ] is the round function. Then, we change color of the pixel $c_{ij}$ with coordinates of $(i = u(x_n), j = u(y_n))$ into the opposite color, i.e., if color of the pixel be white, it changes black and vice versa. In other words, if black and white colors be showed 0 and 1, respectively, those can be changed the following method,

$$c_{ij} = \begin{cases} 0 \to 1 \\ \text{or} \\ 1 \to 0 \end{cases} . \quad (4)$$

We iterate this method (Eq. 3,4) $M$ times. Matrix that is created with this method, we show $C'_{2^k \times 2^k}$. The $M$ value is related to the two tests. So that, we create two bit sequence from the matrix $C'_{2^k \times 2^k}$. The first bit sequence to join the rows of the matrix is formed and the second bit sequence to join the columns of the matrix is formed. If two bit sequences to satisfy Monobit Test and Serial Test (see Appendix) separately, then, the M value is the correct value. Consequently, the resulting black white image (the black white image of the matrix $C'_{2^k \times 2^k}$) is the perfect seed for PRNG (see Fig.1(c)). For generating random bit sequence from this method, we are using a two-dimensional chaotic system which is defined as follows:

$$x'_n = f'(x_{n-1}) \qquad n = 0,1,2,...$$
$$y'_n = g'(y_{n-1}) \qquad n = 0,1,2,...$$

that $f'$ and $g' : I \to I$ ($I = [0,1]$) are nonlinear maps. Thus, using the transformation of Eq.3, the random bit sequence $\{z_n\}_{n=0}^{\infty}$ is defined as follows:

$$z_n = \begin{cases} 0 & c'_{i'j'} = 0 \\ 1 & c'_{i'j'} = 1 \end{cases} \quad (5)$$

and because the black white image be a black white dynamic image, after each iteration of the Eq. 5 with this method adds the following term:

$$c'_{i'j'} = \begin{cases} 0 \to 1 \\ \text{or} \\ 1 \to 0 \end{cases} . \quad (6)$$

Therefore, we get a black white dynamic image as a seed for the proposed PRNG. As an example, we consider the logistic map (1) for the functions of $f$, $f'$, $g$ and $g'$ ($x_n, x'_n, y_n, y'_n \in (0,1)$, and $r_x, r_{x'}, r_y, r_{y'} \in (3.99996, 4]$) (see Fig.2).

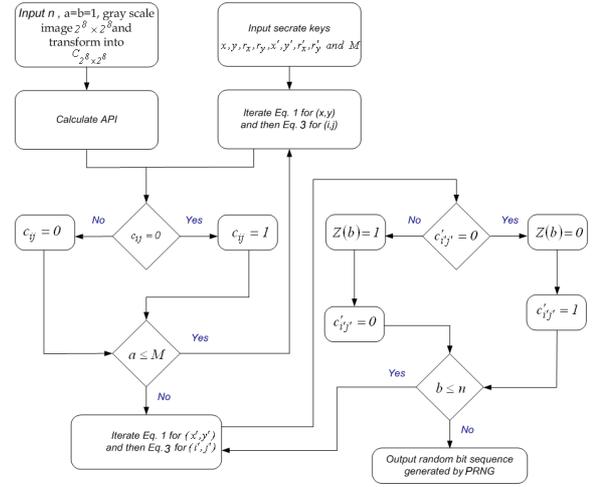

*Fig. 2. Block diagram of the logistic pseudorandom image algorithm for generation pseudorandom bit sequence.*

### 3.1 Analysis of randomness of number sequences

We have survey the randomness and uniformity of the several bit sequences of large size, generated by the proposed PRNG for different sets of control parameter and initial conditions of chaotic logistic maps and images. Here, we show the results for $2^{20}$ sized bit sequences corresponding to the following parameter values of the four sets:

$$\begin{cases} A = (0.2, 0.6, 4, 3.99997, 0.5, 0.1, 3.99998, 3.99999, 2^{18}) \\ B = (0.7, 0.3, 3.99998, 3.99996, 0.8, 0.4, 3.99997, 3.99999, 2^{18}) \\ C = (0.4, 0.8, 3.99996, 4, 0.6, 0.2, 3.99998, 3.99999, 2^{17}) \\ D = (0.7, 0.2, 3.99999, 3.99997, 0.3, 0.6, 3.99998, 4, 2^{17}) \end{cases}$$

For convenience, these four sets are designated as: $\{A, B, C, D = (x, y, r_x, r_y, x', y', r_{x'}, r_{y'}, M)$

that A, B, C and D are related control parameter values of PRNG (see Table.1). We have used MATLAB 7.10.0 (R2010a) running program in a personal computer with a Core i3 3.1GHz intel, 4GB memory and 500GB hard-disk capacity. The average time used for generating random bit sequences with size of $2^{20}$ bits is shorter than 0.4 s.

We discuss in the following paragraph of this Section the result and conclusions of our study of the different statistical tests to observe the randomness and uniformity of the bit sequences generated by the proposed PRNG. The US NIST statistical test suite provides 15 statistical tests to detect deviations of a bit sequence from randomness. A statistical test is formulated to test a null hypothesis which states that the sequence being tested is random. There is also an alternative hypothesis which states that the sequence is not random. For each test, there is an associated reference distribution (typically normal distribution or $\chi^2$ distribution), based on which a P-value is computed from the bit sequence. If the P-value is greater than a predefined threshold $\alpha$ which is also called significance level, then the sequence would be considered to be random with a confidence of $1 - \alpha$, and the sequence passes the test successfully. Otherwise, the sequence fails this test. A P-value of zero indicates that the sequence appears to be completely non-random, and the larger the P-value is, the closer a sequence to a perfect random sequence. In our experiment, we set $\alpha$ to its default value 0.01, which means a sequence passed the test is considered as random with 99% confidence. Before presenting the test results of our proposed three approaches, we would first introduce all 15 statistical tests briefly as follows. A more detailed description for those tests could be found in [2].

Table I. Shows that the M value is the correct value if and only if two bit sequences created of the rows of the matrix $2^8 \times 2^8$ (1) and the columns of the matrix $2^8 \times 2^8$ (2) pass monobit test and serial test.

| Parameter | Calculated $\chi^2$ value | | | | Critical $\chi^2$ value at $\alpha = 0.05$ | |
|---|---|---|---|---|---|---|
| | Monobit test | | serial test | | Monobit test | serial test |
| | (1) | (2) | (1) | (2) | | |
| A (A. Einstein image) | 0.7985 | 1.0172 | 1.8143 | 1.8871 | 3.8415 | 5.9915 |
| B (Tree image) | 1.0378 | 1.1035 | 2.0568 | 2.0270 | 3.8415 | 5.9915 |
| C (A. Einstein image) | 1.0303 | 1.0098 | 2.1246 | 1.9883 | 3.8415 | 5.9915 |
| D (Tree image) | 1.4295 | 1.4487 | 2.4115 | 2.5635 | 3.8415 | 5.9915 |

Table II. Shows the P-values obtained from NIST suite for fifteen different tests. The P-values are obtained for four different sets of parameters for each test.

| NIST Tests | | A (A. Einstein image) | B (Tree image) | C (A. Einstein image) | D (Tree image) |
|---|---|---|---|---|---|
| FT | | 0.979743 | 0.600670 | 0.956387 | 0.284479 |
| FTB | | 0.873583 | 0.794484 | 0.961466 | 0.218437 |
| RT | | 0.863536 | 0.799775 | 0.766560 | 0.047121 |
| LROBT | | 0.953186 | 0.643394 | 0.928064 | 0.287490 |
| BMRT | | 0.920326 | 0.143269 | 0.273873 | 0.649518 |
| DFTT | | 0.372087 | 0.544647 | 0.482314 | 0.214210 |
| NTMT | | SUCCESS | SUCCESS | SUCCESS | SUCCESS |
| OTMT | | 0.665345 | 0.093392 | 0.764690 | 0.399512 |
| MUST | | 0.971350 | 0.165435 | 0.278815 | 0.218812 |
| LCT | | 0.869026 | 0.424203 | 0.246919 | 0.597068 |
| ST | P1 | 0.176425 | 0.807509 | 0.038659 | 0.155790 |
| | P2 | 0.062528 | 0.867147 | 0.108128 | 0.355935 |
| AET | | 0.198495 | 0.905032 | 0.548792 | 0.166571 |
| CST | (FORWARD) | 0.999421 | 0.982586 | 0.977552 | 0.460996 |
| | (REVERSE) | 0.998589 | 0.861198 | 0.991191 | 0.556137 |
| RET | | SUCCESS | SUCCESS | SUCCESS | SUCCESS |
| REVT | | SUCCESS | SUCCESS | SUCCESS | SUCCESS |

The frequency test (FT), the runs test (RT) and the cumulative sum test (CST) are recommended that each sequence to be tested consist of a minimum of $10^2$ bits (i.e., $n \geq 10^2$). The frequency Test within a Block (FTB) is recommended that each sequence to be tested consist of a minimum of $M \times N$ bits (i.e., $n \geq MN$). The block size $M$ should be selected such that $M \geq 20$ and $N < 10^2$. The discrete fourier transform test (DFTT) is recommended that each sequence to be tested consist of a minimum of $10^3$ bits (i.e., $n \geq 10^3$). The approximate entropy test (AET) is recommended that each sequence to be tested consist of a minimum of $2^{12}$ bits (i.e., $n \geq 2^{12}$). The test for the longest run of ones in a block (LROBT) is recommended that each sequence to be tested consist of a minimum of 6272 bits for M=128. The binary matrix rank test (BMRT) is recommended that each sequence to be tested consist of a minimum of $10^5$ bits (i.e., $n \geq 10^5$). The non-overlapping template matching test (NTMT), the overlapping template matching test (OTMT), the maurer's universal statistical test (MUST), the linear complexity test (LCT), the serial test (ST), the random excursions test (RET) and the random excursions variant test (REVT) are recommended that each sequence to be tested consist of a minimum of $2^{20}$ bits (i.e., $n \geq 2^{20}$).

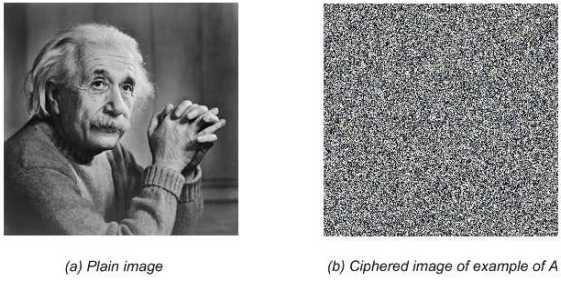

*Fig. 3. Images of test results.*

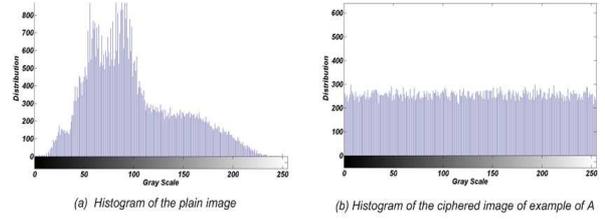

*Fig. 4. Histograms of images.*

The NIST suite tests were performed on four bit sequences, each containing $2^{20}$ bits. The P-value as well as final results obtained from the NIST suite for four different sets are given in Table 2. The proposed PRNG successfully passes all randomness tests of NIST suite.

## 4. The proposed encryption scheme and security analysis

### 4.1 Encryption scheme based on pseudorandom bit padding

In the proposed scheme, we create a method to encrypt the image using bits padding. To consider a gray scale image with the size of $M \times N$. Here, the plain image is the image of the example of A that an image with the size of $256 \times 256$ (see Fig. 3(a)). The steps of the encryption are shown below:

- Step 1: Generate $8 \times M \times N$ pseudo-random number sequence using the logistic pseudorandom image algorithm.
- Step 2: Transform the image into $8 \times M \times N$ bit sequence (image sequence).
- Step 3: Perform the XOR operation between the image sequence and the pseudo-random bit sequence to form the cipher sequence.
- Step 4: Transform the cipher sequence into image matrix $I$ (ciphered image).
- Step 5: Divide the matrix $I$ into four parts, uniformly. Move the odd rows with the even rows between the two parts in the main diagonal and between the other two parts, respectively.
- Step 6: Divide the matrix $I$ into four parts, uniformly. Move the odd columns with the even columns between the two parts in the main diagonal and between the other two parts, respectively.

The ciphered image is shown in fig. 3(b). The grey scale histograms are given in figs. 4(a), 4(b). The fig. 4(b) shows uniformity in distribution of grey scale of the ciphered images. In addition, the average pixel intensity for plain image is 98.92 and for ciphered image is 127.09.

### 4.2 Analysis of security of the proposed encryption scheme

The Security is a major intransitive of a cryptosystem. Here, a complete analysis is made on the security of the cryptosystem. We have tried to explain that this cipher image is sufficiently secure against various cryptographical attacks, as shown below:

#### 4.2.1 Key space analysis

Key space size is the total number of different keys that can be used in the encryption [20]. Security issue is the size of the key space. If it is not large enough, the attackers may guess the image with brute-force attack. If the precision is $10^{-14}$, the size of key space for initial conditions and control parameters is $2^{306}$. In addition, we use the black white dynamic images derived of Albert Einstein image with $256 \times 256$ pixels. The size of the key space for black white dynamic image is no less than $2^{256}$. This size is large enough to defeat brute-force by any super computer today.

#### 4.2.2 Correlation Coefficient Analysis

The statistical analysis has been performed on the encrypted image from example of A. This is shown by a test of the correlation between two adjacent pixels in plain image and encrypted image. We randomly select 2000 pairs of two-adjacent pixels (in vertical, horizontal, and diagonal direction) from plain images and encrypted images, and calculate the correlation coefficients [19,20], respectively by using the following two equations (see Table 3 and Fig. 5(a) and 5(b)):

$$Cov(x,y) = \frac{1}{N}\sum_{i=1}^{N}(x_i - E(x))(y_i - E(y)),$$

$$r_{xy} = \frac{Cov(x,y)}{(D(x))^{\frac{1}{2}}(D(y))^{\frac{1}{2}}}$$

Where

$$E(x) = \frac{1}{N}\sum_{i=1}^{N}(x_i), \quad D(y) = \frac{1}{N}\sum_{i=1}^{N}(x_i - E(x))^2.$$

*Where, E(x) is the estimation of mathematical expectations of x, D(x) is the estimation of variance of x, and Cov(x,y) is the estimation of covariance between x and y, where x and y are grey scale values of two adjacent pixels in the image.*

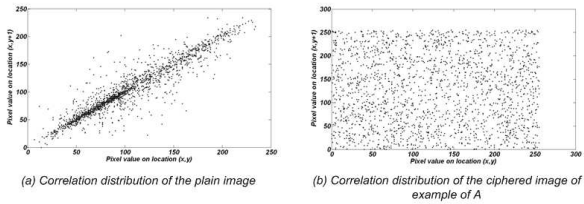

*Fig. 5. Correlation distributions of two horizontally adjacent pixels in the plain image and the ciphered image.*

*Table III. Correlation coefficients of two adjacent pixels in the plain image and the ciphered image of example of A.*

| Direction | Plain image | ciphered image |
|---|---|---|
| Horizontal | 0.9341 | 0.0023 |
| Vertical | 0.9634 | 0.0098 |
| Diagonal | 0.9402 | 0.0043 |

### 4.2.3 Differential attack

Attackers try to find out a relationship between the plain image and the cipher image, by studying how differences in an input can affect the resultant difference at the output in an attempt to derive the key [31]. Trying to make a slight change such as modifying one pixel of the plain image, attacker observes the change of the cipher image [31]. To test the influence of one pixel change on the whole encrypted image by the proposed scheme, two common measures are used:

Number of Pixels Change Rate (NPCR) stands for the number of pixels change rate while, one pixel of plain image is changed. Unified Average Changing Intensity (UACI) measures the average intensity of differences between the plain image and ciphered image. The NPCR and The UACI, are used to test the influence of one pixel change on the whole image encrypted by the proposed scheme and can be defined as following:

$$NPCR = \frac{\sum_{i,j} D(i,j)}{W \times H} \times 100\%$$

$$UACI = \frac{1}{W \times H}\left[\sum_{i,j}\frac{C_1(i,j) - C_2(i,j)}{255}\right] \times 100\%$$

where $W$ and $H$ are the width and height of $C_1$ or $C_2$. $C_1$ and $C_2$ are two ciphered images, whose corresponding original images have only one pixel difference and also have the same size. The $C_1(i,j)$ and $C_2(i,j)$ are grey-scale values of the pixels at grid $(i,j)$. The $D(i,j)$ determined by $C_1(i,j)$ and $C_2(i,j)$. If $C_1(i,j) = C_2(i,j)$, then, $D(i, j) = 1$; otherwise, $D(i, j) = 0$. We have done some tests on the proposed scheme (256 grey scale image of size $256 \times 256$) to find out the extent of change produced by one pixel change in the plain image. We have obtained NPCR = 0.43% and UACI = 0.34%. The results demonstrate that the proposed scheme can survive differential attack.

## 5. Conclusion

We have proposed a chaotic encryption scheme based on pseudorandom bit padding that the bits be generated by a novel logistic pseudorandom image algorithm. The security of the cipher image of this scheme is evaluated by the key space analysis, the correlation of two adjacent pixels and differential attack. The distribution of the ciphered images is very close to the uniform distribution, which can well protect the information of the image to withstand the statistical attack.

*Appendix*

*Monobit Test:*
The goal of this test is to determine whether the frequency of 0's and 1's in bit sequences generated by the PRNG are approximately same [27]. Let $n_0, n_1$ denote the number of 0's and 1's in bit sequences respectively. We calculate $\chi^2$ by using the formula [27]:

$$\chi^2 = \frac{(n_0 - n_1)^2}{n},$$

which approximately follow a $\chi^2$ distribution with one degree of freedom. The computed results are shown in Table 1. The calculated values of $\chi^2$ are less in compared to the critical value of $\chi^2$ at $\alpha = 0.05$ (5% level of significance) and 1df (one degree of freedom). It means that these bit sequences pass the monobit test and can be said to be satisfactorily random with respect to this test [27].

*Serial Test:*

The goal of this test is to determine whether the number of occurrence of pairs 00, 01, 10 and 11 in the bit streams generated by PRNG is approximately same [27]. Let $n_{00}, n_{01}, n_{10}$ and $n_{11}$ denote the number of occurrence of pairs 00, 01, 10 and 11 respectively in the bit sequences. We calculate $\chi^2$ by using the formula [27]:

$$\chi^2 = \frac{4}{n-1}(n_{00}^2 + n_{01}^2 + n_{10}^2 + n_{11}^2) - \frac{2}{n}(n_0^2 + n_1^2) + 1$$

and the computed values are found to follow approximately the $\chi^2$ distribution with 2 degrees of freedom. The results are shown in Table 2. The calculated values of $\chi^2$ are less than critical value of $\chi^2$ at $\alpha = 0.05$ (5% level of significance) and 2df (two degrees of freedom). It means that bit sequences pass the serial test and are satisfactorily random with respect to this test.